\documentstyle[12pt,epsfig]{article}
\begin{document}
\topmargin -0.5cm
\oddsidemargin -0.3cm

\begin{titlepage}
\pagestyle{empty}
\begin{flushright}
{CERN-TH.7217/94}
\end{flushright}
\vspace*{1mm}
\begin{center}
{\bf FIRST EVIDENCE FOR ELECTROWEAK}\
{\bf RADIATIVE CORRECTIONS FROM THE NEW }\
{\bf PRECISION DATA}\\
\vspace{0.3in}
{\bf  V.A. Novikov}$^{*)}$ ,\\
\vspace{0.1in}
University of Guelph, Guelph, ON, N1G2W1,\\
Canada\\
{\bf L.B.  Okun}$^{*)}$ ,\\
\vspace{0.1in}
Theoretical Physics Division, CERN\\
CH-1211 Geneva 23, Switzerland \\
\vspace{0.1in}
and \\
{\bf A.N. Rozanov}$^{**)}$,
{\bf M.I. Vysotsky}\\
\vspace{0.1in}
ITEP, Moscow, 117259, Russia\\
\vspace{0.5in}
 {\bf Abstract \\}
 \end{center}
The analysis of the newest data on the leptonic $Z$-decays
 and $m_W$ appears to reveal the first
manifestations of electroweak radiative corrections.
 In fact, these data differ,  at the level of
$2\sigma$, from their electroweak Born values,
while they agree, to within $1\sigma$, with the theoretical values
which take  the electroweak radiative corrections into account.
Previous data were within $1\sigma$ in agreement with
both sets of values.

\vspace{4cm}
\noindent
\rule[.1in]{14.0cm}{.002in}

\noindent
$^{\, *)}$ Permanent address: ITEP, Moscow 117259, Russia. \\
$^{**)}$ Present address:
Particle Physics Experiments Division, CERN\\
CH-1211  Geneva 23, Switzerland \\

\begin{flushleft}
CERN-TH.7217/94 \\
April 1994
\end{flushleft}
\end{titlepage}

\vfill\eject
\pagestyle{empty}

\setcounter{page}{1}
\pagestyle{plain}

The traditional way of analyzing the data on electroweak radiative
 corrections, (see
for instance \cite{A} - \cite{C}), is to $not$ split off from them
the large and purely electromagnetic effect of the running of  the
electric charge from $q^2 = 0$ to $q^2 = m_Z^2$.
According to that approach, which starts from
\linebreak
 $\alpha \equiv \alpha(0) = 1/ 137.0359895(61)$,
the ``electroweak" corrections appear to be large and to have been
 observed for a long time.
By analyzing them, many authors \cite{topold} came already
 several years ago
 to the conclusion that the mass
 of the top quark must be close to 130 GeV or heavier.

 In  a series of papers
  \cite{th6053}-\cite{5} we developed an approach in which
the running of $\alpha(q^2)$ is
explicitly excluded from the genuinely electroweak corrections
 and included in
the electromagnetic ones.
Our main argument is that the running of $\alpha(q^2)$
up to $q^2 = m^2_Z$ is a purely
electromagnetic phenomenon which is totally insensitive to the
existence of electroweak bosons (W, Z and higgs), and that
 $\alpha(0)$,
with all its impressive accuracy, is wholly irrelevant to electroweak physics
even at low energy \cite{th7153}.
Our approach starts with the most accurately known
electroweak observables:
\begin{equation}
G_{\mu}= 1.16639(2)\cdot 10^{-5} \hspace{3mm}
\mbox{\rm GeV}^{-2}~, \hspace{2cm} \cite{rpd}
\label{1}
\end{equation}
\begin{equation}
m_Z = 91.1899(44) \hspace{3mm}
 \mbox{\rm GeV}~,  \hspace{2.4cm} \cite{6}
\label{2}
\end{equation}
\begin{equation}
\bar{\alpha}\equiv \alpha(m_Z) = 1/128.87(12)~, \hspace{2.1cm}
  \cite{jegerlehner}
\label{3}
\end{equation}
and has three free parameters: the top quark mass, $m_t$, the Higgs
 boson mass,
$m_H$, and the QCD coupling constant
 $\bar{\alpha}_s \equiv \alpha_s(m_Z)$.
The conventional nature of the definition on $\bar{\alpha}$ is analyzed in
\cite{th7071}.

In terms of $G_{\mu}, m_Z$ and $\bar{\alpha}$ we define the electroweak
angle $\theta$ ($sin\theta \equiv s, cos\theta \equiv c$)
\cite{th6053}, \cite{D}, \cite{peskin}:
\begin{equation}
s^2 c^2 = \frac{\pi \bar{\alpha}}{\sqrt{2} G_{\mu} m_Z^2},
\label{4}
\end{equation}
which is analogous to, but different from, the traditional $\theta_W$
($sin \theta_W \equiv s_W,$
$cos \theta_W \equiv c_W)$ defined by substituting  $\alpha$ instead of
 $\bar{\alpha}$ in eq.(\ref{4}).
By solving eq.(\ref{4}) one finds:
\begin{equation}
s^2 = 0.23118(33), \hspace{2cm}   c = 0.87682(19)
\label{5}
\end{equation}
In the $\bar{\alpha}$-Born approximation
\begin{equation}
m_W/m_Z = c = 0.8768(2),
\label{6}
\end{equation}
\begin{equation}
g_A = - 1/2,
\label{7}
\end{equation}
\begin{equation}
g_V/g_A = 1 - 4s^2 = 0.0753(12).
\label{8}
\end{equation}
Here $g_V$ and $g_A$ are the vector and axial couplings of the
Z boson decay into a pair of charged leptons $l\bar{l}$.
(Note that with the traditional angle $\theta_W$ we would get
$s^2_W = 0.2122$ and in the $\bar{\alpha}$-Born approximation
$g_V/g_A = 0.1514$ which differs by $40 \sigma$ (!)   from the
corresponding experimental value (see Table 1).

The width of the decay $Z \rightarrow l\bar{l}$ is given by expression:
\begin{equation}
\Gamma_l = 4 (1+\frac{3\bar{\alpha}}{4 \pi})(g^2_A + g^2_V) \Gamma_0,
\label{9}
\end{equation}
where
\begin{equation}
\Gamma_0 = \frac{\sqrt{2}G_{\mu}m^3_Z}{48\pi} = 82.948(12)
\hspace{5mm} \mbox{\rm MeV}
\label{10}
\end{equation}
The first bracket in eq. (\ref{9}) takes into account the purely
 electromagnetic
corrections.

In a similar manner, the width of Z decaying into a pair of quarks
$q \bar{q}$ with charge $Q$ and the isospin projection $T_3$
 is given by
\begin{equation}
\Gamma_q = 12 (1 +\frac{3 Q^2\bar{\alpha}}{4\pi})
(g^2_{Aq} + g^2_{Vq}) \Gamma_0 G
\label{11}
\end{equation}
where
\begin{equation}
g_{Aq} = T_3,
\label{12}
\end{equation}
\begin{equation}
g_{Vq} / g_{Aq} = 1 - 4 |Q| s^2.
\label{13}
\end{equation}
The extra factor of 3, as compared with eq.(\ref{9}),
comes from the colour
 and the factor $G$ takes into
account the emission and exchange of gluons \cite{kataev}:
\begin{equation}
G = 1 + \bar{\alpha_s}/ \pi
      + 1.4 (\bar{\alpha_s}/ \pi)^2
      - 13  (\bar{\alpha_s}/ \pi)^3  + ...
\label{14}
\end{equation}

We thus define the $\bar{\alpha}$-Born approximation for $\Gamma_l$
by eqs.(\ref{7})-(\ref{10}) and for $\Gamma_h$ by  summing
  eq. (\ref{11}) over all quarks, thereby taking into account the
 QED and QCD loop corrections.
 Beyond the $\bar{\alpha}$-Born approximation, one has to include in
 $g_A, g_V, g_{Aq}, g_{Vq}$ the contributions of electroweak loops
 proportional to $\bar{\alpha}/ \pi$
 (with gluonic corrections in some of them).

In ref. \cite{1} we concluded that the data of four LEP detectors, announced
at the 1993 La Thuile \cite{2} and Moriond \cite{3} conferences, were,
within $1\sigma$, described by the electroweak $\bar{\alpha}$-Born
approximation as well as by the standard model expressions
including the one-loop electroweak corrections. This means that the genuine
electroweak corrections were not visible
experimentally at that time.

The non-observation of deviations from the electroweak $\bar{\alpha}$-Born
approximation, with due allowance for QED and QCD effects,
 enabled us to predict
  the values of
$\bar{\alpha}_s$ and $m_t$
 within the framework of
the Minimal Standard Model,
while $m_H$ remained practically
non-constrained. In this respect our results did not differ from those of
the traditional  approach. In our approach the possibility of constraining
 $m_t$ arises from the mutual compensation of the contributions of the
top quark and all other virtual particles for $m_t$ in the range of $160\pm
20$ GeV \cite{1}.

The experimental data  changed somewhat by the time of the Marseille
Conference \cite{4},\cite{C},
 so that the maximal deviation from the corresponding
$\bar{\alpha}$-Born value became $1.3 \sigma$ (for $g_V/g_A$) \cite{5}.
Obviously,
 the situation did not change qualitatively.

According to the fit of ref. \cite{5}, the values of the LEP observables
were equally well
 described within $1\sigma$ by the $\bar{\alpha}$-Born approximation and
  by the Minimal Standard Model amplitudes
including the electroweak radiative corrections. The only exception was the
value of $R_b$ for a heavy higgs where discrepancy
 with the MSM prediction
 reached $1.7 \sigma$.
(See Table 1 from \cite{5}.)

At the
1994 La Thuile and Moriond conferences \cite{6}
 new, more accurate data were  presented by CDF, ADLO and SLD.
In the present note we compare
these data with our theoretical expressions, which have been combined into a
computer code called LEPTOP
\footnote{One can obtain the FORTRAN code of LEPTOP from
rozanov@cernvm.cern.ch}.

 Let us start by considering the data of CDF and ADLO.
{}From  Table 1 we see that the new experimental values of $m_W/m_Z$,
$\Gamma_l$ and $g_V/g_A$ deviate from their $\bar{\alpha}$-Born value
by $2\sigma$. These are the so-called ``gluon-free" observables \cite{novy93}
which depend on $\bar{\alpha_s}$ only very weakly, i.e., only through terms of
the order of $\bar{\alpha}\bar{\alpha_s}$.
 At the same time the data agree within $1\sigma$ with those
theoretical predictions which take the electroweak radiative
corrections into account.
{\it We consider this as a first indication that the genuine
electroweak corrections have become observable.}
This conclusion is strengthened by the fact that the experimental errors in
$m_W/m_Z$, $\Gamma_l$ and $g_V/g_A$ are practically uncorrelated.
  Note the difference
between our statement and that of Ref. \cite{7} where the departure of
 the MSM predicted (fitted) values from the $\bar\alpha$-Born ones is being
stressed.

There are two small clouds on this blue sky.
First, the new measurements of $A_{LR}$ at SLD give
$sin^2\theta_{eff}=0.2290(10)$ or $g_V/g_A = 0.0840(40)$, which
differs by $3\sigma$
 from the LEP value $g_V/g_A = 0.0711(20)$ and from the
theoretical
prediction (see Table 1).
This discrepancy is probably of purely experimental origin. Note that the SLD
value for $g_V/g_A$ lies $2\sigma$ above the $\bar\alpha$-Born value, while the
LEP value lies  $2\sigma$ below. Their average is compatible with
$\bar\alpha$-Born.

Second, the value of $R_b$ measured at LEP coincides with the
 $\bar{\alpha}$-Born
value and
 is $2.5\sigma$ away from its
theoretically fitted value $R_b = 0.2161(4)^{-6}_{+6}$ with
the
central value corresponding to $m_H = $ 300 GeV, the shifts + (--) 6 to $m_H$ =
60(1000) GeV, and the uncertainty $\pm 4$ to
 $\delta m_t = \pm 11$ GeV.
 This discrepancy may, if  not caused by a systematic
error,  indicate the existence of new physics \cite{4}.

Let us note that the figures presented in the Table correspond to the fitted
values of $m_t$ and $\bar{\alpha_s}$
derived from the new LEP and CDF data:
\begin{equation}
m_t = 171(11)^{+15}_{-21}(5),
\label{15}
\end{equation}
\begin{equation}
\bar{\alpha}_s \equiv \alpha_s(m_Z) = 0.125\pm 0.005 \pm 0.002,
\label{16}
\end{equation}
\begin{equation}
\chi^2 = 14/10.
\label{17}
\end{equation}
Here the central values correspond again to $m_H=300$ GeV, with
the first uncertainties  being experimental, the second corresponding to $m_H =
300^{+700}_{-240}$ GeV, and the third (for $m_t$) corresponding to the
uncertainty in $1/\bar{\alpha} = 128.87 \pm 0.12$.

Comparing this with the fit
  \cite{5}
 of the earlier data:
\begin{equation}
m_t = 162^{+14 +16}_{-15 -22},
\label{18}
\end{equation}
\begin{equation}
\bar{\alpha}_s = 0.119 \pm 0.006 ^{+0.002}_{-0.003},
\label{19}
\end{equation}
\begin{equation}
\chi^2 = 3.5/10,
\label{20}
\end{equation}
we observe that central values of $m_t$ and $\alpha_s$ have increased, their
uncertainties decreased, while the $\chi^2$  became more palatable.
The individual contributions to the average value of $m_t$ show more
variations than previously (see Fig. 1).

Our new fitted values for $m_t$ and $\bar{\alpha_s}$ are
in good agreement with these of the LEP
Electroweak Working Group as obtained in the traditional approach and
 presented at the Moriond Conference \cite{6}.

The numbers of the fit (\ref{15})--(\ref{17}) and of Table 1  include a
recently estimated QCD
correction
 \cite{8},
which increases $m_t$ by about 4 GeV.

With reference to Table 1, we would like to stress two points:
\begin{itemize}
\item[(1)]
The shifts caused by changing $m_H$ are, as a rule, small compared to the
uncertainties (in brackets) in column 5. This ``$m_H$ independence" is
characteristic for the global fit which predicts $m_t$ for a given $m_H$. The
higher $m_H$, the higher is the predicted $m_t$, while the predicted values of
the observables remain practically unchanged. (This would be evident if there
was only a single observable).
\item[(2)] The situation is different when $m_t$ is fixed (e.g., measured). For
$m_t =$ 170 GeV, the shifts of $g_V/g_A$ from its central value 0.0711 are
--0.0024 and +0.0035 for $m_H$ = 1000 GeV and 60 GeV,
 respectively (see Table 2 of
Ref. [6]), which is larger than the current experimental
uncertainty in $g_V/g_A (\pm$ 0.0020). Thus a further improvement of
 the accuracy
in $g_V/g_A$ could place serious bounds on $m_H$. Two other ``gluon-free"
observables, $m_W/m_Z$ and $g_A$, are less sensitive:
 their higgs shifts are half
as large as their present experimental uncertainties.
\end{itemize}

To conclude: Within the framework of the traditional approach,
 which starts with
$\alpha(0)$, the latest  precision data
do not herald anything qualitatively new;
one merely gets a slightly heavier  top  mass, and a
slightly larger  strong coupling constant.
In strong contrast, these same data open,
 with our approach -- which starts with
$\alpha (m_Z)$ -- a new window, one through which
 the non-vanishing electroweak
radiative corrections become visible.

\vspace*{1cm}
\noindent
{\rm ACKNOWLEDGEMENTS}

We are grateful to D.Yu.Bardin, A.Sirlin, V.L.Telegdi and M.B.Voloshin
 for helpful remarks.
VN, LO, and MV are grateful to the Russian Foundation for Fundamental
Research for grant 93-02-14431. LO, MV and AR
 are grateful to CERN TH and PPE
Divisions, respectively, for their warm hospitality.

\newpage
\begin{center}

{\large Table 1}
\end{center}

   Results of fitting the
    Moriond 1994 data from LEP and $p \bar p$ colliders.
   Observables (first column), their '94 and '93 experimental values (second
and third columns) and
   their predicted values: (a) in the electroweak tree (Born)
    approximation based
on
   $\bar{\alpha}$ (fourth column) and (b) in the electroweak
    tree plus one loop
   approximation (fifth column). Both in columns 4 and 5 the QED and QCD
   loops were taken into account.

   The predicted values have been obtained for three fixed values
   of $m_H = 300^{+700}_{-240}$ GeV;
   for each of them the fitted values of $m_t \pm \delta m_t$
   and $\bar{\alpha_s} \pm \delta \alpha_s$ were used.
   The central values correspond to $m_H=300$ GeV. The upper (lower)
   numbers give the shifts of these central values corresponding to
   $m_H = 1000$  (60) GeV.

   The numbers in brackets correspond to experimental uncertainties (columns 2
and 3),
   and  predicted uncertainties (columns 4 and 5), arising
   in column 4 from $\delta \bar{\alpha}$ for $m_W/m_Z$,
    $g_V/g_A$ and $\Gamma_l$ and from $\delta \bar{\alpha}_s$
    for the five other observables.
    The errors in brackets in column 5 come from
    $\delta \bar{\alpha_s}$ and $\delta m_t$ of the fit and from
    $\delta \bar{\alpha}$ (for $g_V/g_A$ only).
    Note that the $\bar{\alpha}$-Born values of hadronic observables
    depend on $m_H$. This is caused by their dependence on
    $\bar{\alpha_s}$, the fitted values of which depend on $m_H$.

\vspace{8mm}
\begin{tabular}{|l|l|l|l|l|} \hline
Observable & Exp. '94 & Exp. '93 & $\bar{\alpha}$-Born
 & MSM prediction
 \\ \hline
$m_W/m_Z$ & 0.8814(21)  & 0.8798(28) & 0.8768(2) & 0.8803$(8)^{+0}_{-2}$
 \\ \hline
$g_V/g_A$ & 0.0711(20)  & 0.0716(28) & 0.0753(12) & 0.0711$(19)^{-7}_{+9}$
 \\ \hline
$\Gamma_l$ (MeV) & 83.98(18) & 83.82(27)  & 83.57(2) & 83.87$(11)^{+0}_{-6}$
 \\ \hline
$\Gamma_h$ (GeV) & 1.7460(40) & 1.7403(59) & 1.7445$(26)^{+11}_{-9}$ &
1.7435$(27)^{-3}_{-5}$
 \\ \hline
$\Gamma_Z$ (GeV) & 2.4971(38) & 2.4890(70) & 2.4930$(26)^{+10}_{-10}$ &
 2.4962$(32)^{-3}_{-12}$
 \\ \hline
$\sigma_{had}$ (nb) &   41.51(12) & 41.56(14) & 41.41$(3)^{-10}_{+9}$ &
 41.43$(3)^{+0.2}_{-0.6}$
 \\ \hline
$R_l$ & 20.790(40)  & 20.763(49) & 20.874$(31)^{+13}_{-11}$
  &  20.788$(32)^{-5}_{+10}$
 \\ \hline
$R_b$ & 0.2210(19) & 0.2200(27)  & 0.2197$(0)^{+0}_{-0}$
 & 0.2161$(4)^{-6}_{+6}$
  \\ \hline
\end{tabular}

\newpage
\vspace*{1cm}
\noindent
{\bf Figure  Captions}

\noindent
{\bf Fig. 1}: The fitted values of $m_t$ from the specified observables
measured
at LEP and $p\bar{p}$ colliders, assuming $m_H = 300$ GeV and
$\bar{\alpha}_s = 0.125$. The  region   $m_t < m_Z$,
 is definitely excluded by the direct searches. The
central values of $m_t$ from $R_b$, $A^e_{\tau}$ and $R_l$
 lie in this excluded region.

\noindent
{\bf Fig. 2}: Allowed region of $m_t$ and $m_H$
with $\bar{\alpha_s} = 0.125$. The lines represent the
$s$-standard "ellipses" ($s$=1,2,3,4,5) corresponding
to the constant values of $\chi^2$
($\chi^2 = \chi^2_{min} + s^2$).

\newpage

\begin{figure*}[htb]\centering
\mbox{\epsfig{figure=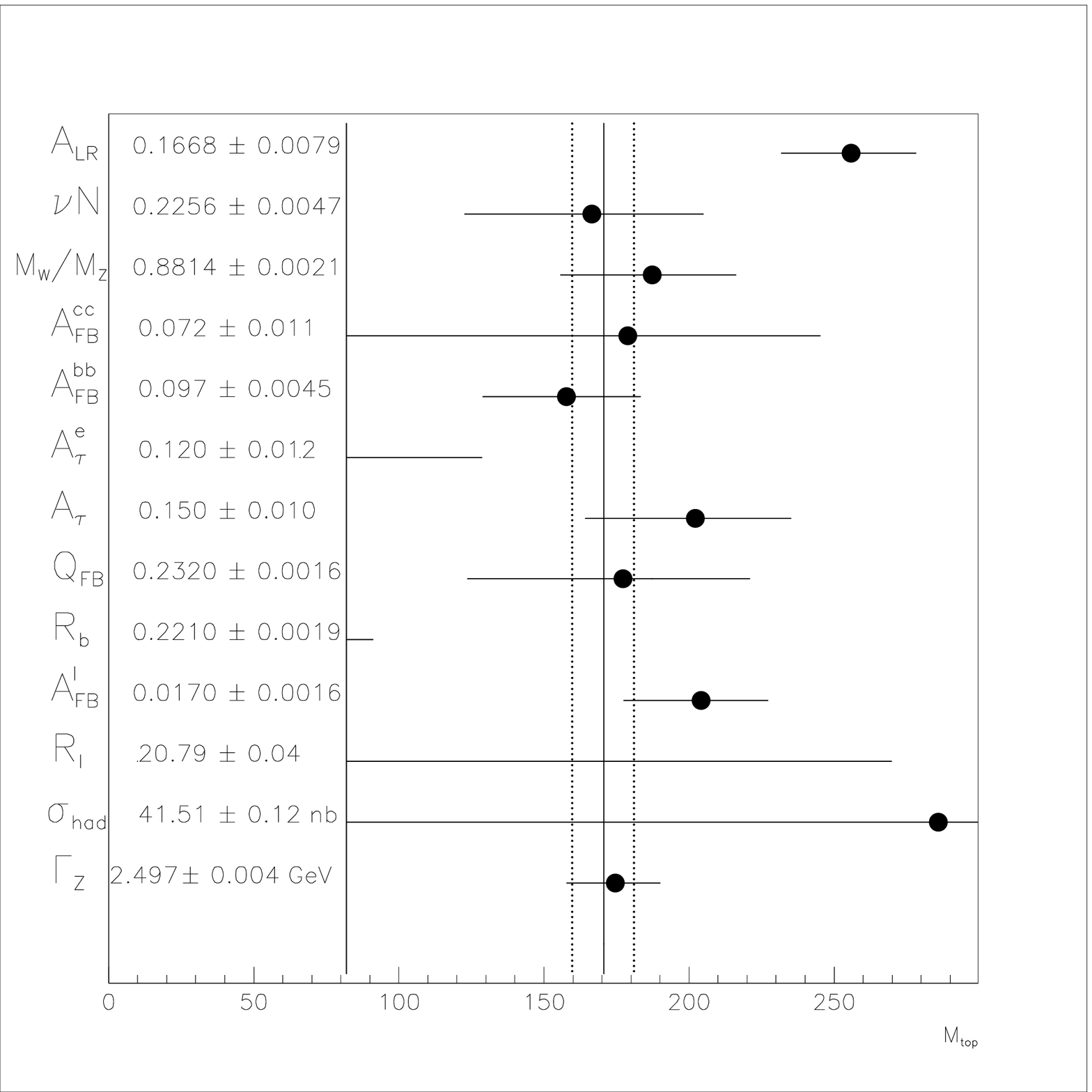,width=0.94\textwidth}}
\caption[]{\label{fig1}
\rm

}
\end{figure*}

\newpage

\begin{figure*}[htb]\centering
\mbox{\epsfig{figure=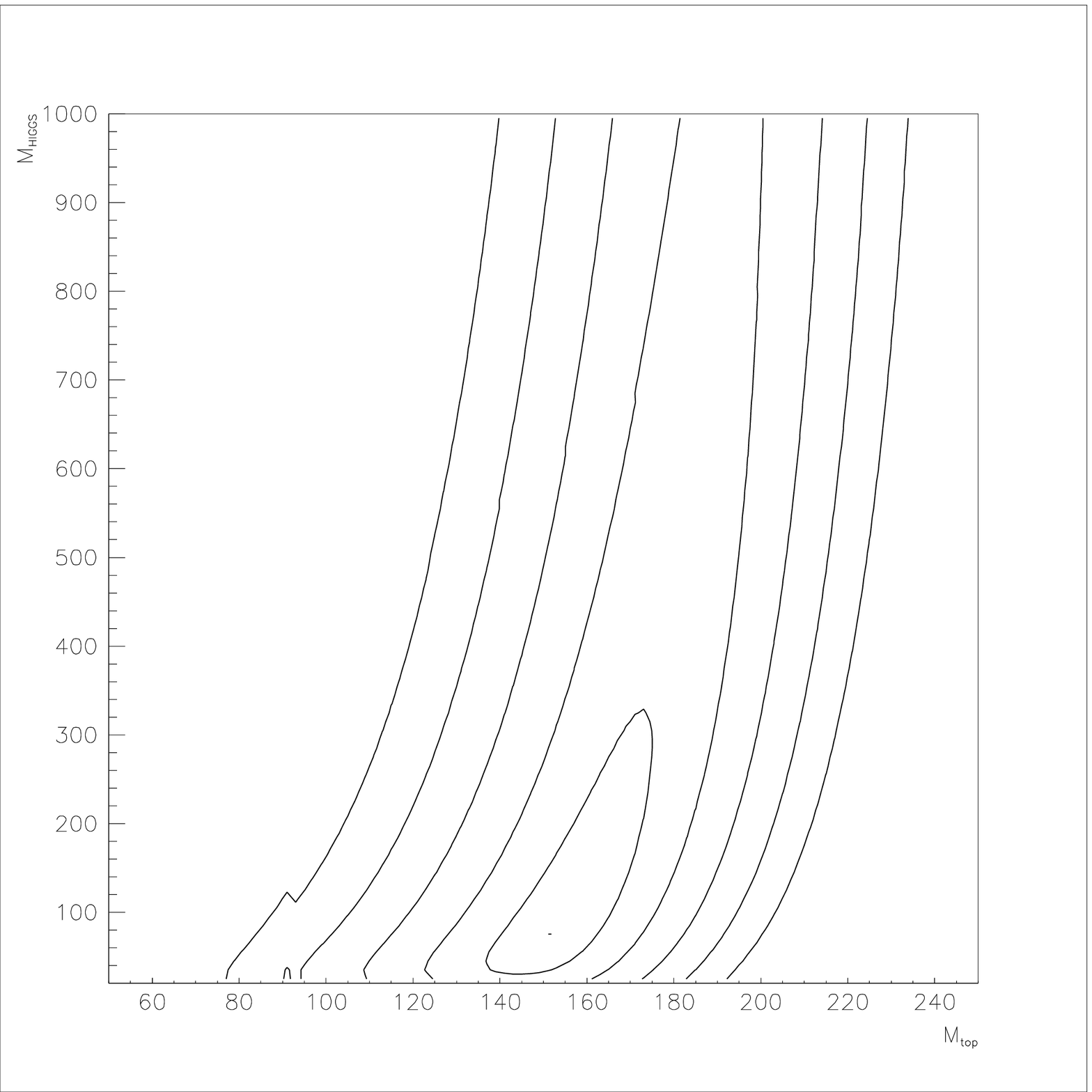,width=0.94\textwidth}}
\caption[]{\label{fig2}
\rm

}
\end{figure*}

\end{document}